\pdfoutput=1

\documentclass[11pt]{article}

\usepackage[final]{acl}

\usepackage{times}
\usepackage{latexsym}

\usepackage[T1]{fontenc}

\usepackage[utf8]{inputenc}

\usepackage{microtype}

\usepackage{inconsolata}

\usepackage{float}

\usepackage{graphicx}
\usepackage{hyperref}
\usepackage{colortbl}
\usepackage{amsmath}
\usepackage{booktabs}
\usepackage{colortbl}

\usepackage{amssymb}
\usepackage{amsmath}
\usepackage{amssymb}
\usepackage{algorithm}
\usepackage{algpseudocode}

\usepackage{pifont}
\let\oldding\ding

\renewcommand{\ding}[2][1]{\scalebox{#1}{\oldding{#2}}}

\usepackage{siunitx}
\usepackage{multirow}

\usepackage{svg}
\usepackage{amsmath}

\usepackage{tikz}
\usepackage{amsmath}
\usepackage{lipsum}

\newsavebox{\myparagraph}
\newlength{\myparheight}
\newlength{\redlinewidth}
\usepackage{fontawesome5}
\usepackage{xcolor}
\usepackage{mdframed} 

\usepackage{authblk}
\usepackage{etoolbox}
\usepackage{marvosym}

\definecolor{lim-red}{RGB}{200,45,55}  
\definecolor{mot-blue}{RGB}{0,90,180}  
\definecolor{code-purple}{RGB}{120,75,155} 
\definecolor{arch-green}{RGB}{0,125,100}

\definecolor{bench-orange}{RGB}{220,110,30}    
\definecolor{model-purple}{RGB}{125,80,155}    
\definecolor{exp-green}{RGB}{0,140,90}

\usepackage{mathtools}
\definecolor{mathblue}{RGB}{0,85,170}

\title{Beyond Function-Level Search: Repository-Aware Dual-Encoder Code Retrieval with Adversarial Verification}

\author[1,2]{Aofan Liu}
\author[1]{Shiyuan Song} 
\author[3]{Haoxuan Li}
\author[1]{Cehao Yang}
\author[1]{Yiyan Qi \Letter}
\affil[1]{\footnotesize International Digital Economy
Academy (IDEA)}
\affil[2]{\footnotesize School of Electronic and Computer Engineering, Peking University}
\affil[3]{\footnotesize Shenzhen International Graduate School, Tsinghua University}

\begin{document}
\maketitle

\begin{abstract}
The escalating complexity of modern codebases has intensified the need for code retrieval systems capable of interpreting cross-component change intents—a capability fundamentally absent in conventional function-level search paradigms. While recent research has improved alignment between queries and code snippets, retrieving contextually relevant code for certain \textit{change request} remains underexplored. \ding[1.2]{182} To bridge this gap, we present \textbf{RepoAlign-Bench}, the first benchmark designed to evaluate repository-level code retrieval for change request-driven scenarios, encompassing 52k columns. The benchmark shifts the paradigm from function-centric retrieval to holistic repository analysis. \ding[1.2]{183} In addition, we propose \textbf{ReflectCode}, an adversarial reflection-augmented dual-tower architecture featuring disentangled \texttt{code\_encoder} and \texttt{doc\_encoder} towers. Our framework dynamically integrates syntactic patterns, function dependency, and semantic expansion intent through LLM. \ding[1.2]{184} Comprehensive evaluations demonstrate that ReflectCode achieves 12.2\% Top-5 Accuracy and 7.1\% Recall improvements over state-of-the-art baselines. Our dataset is available at: \href{https://huggingface.co/datasets/bPtBvkTP/RepoAlignBench}{RepoAlignBench-Full}
\end{abstract}

\section{Introduction}

\begin{figure*}[htbp]
    \centering
    \includegraphics[width=\textwidth]{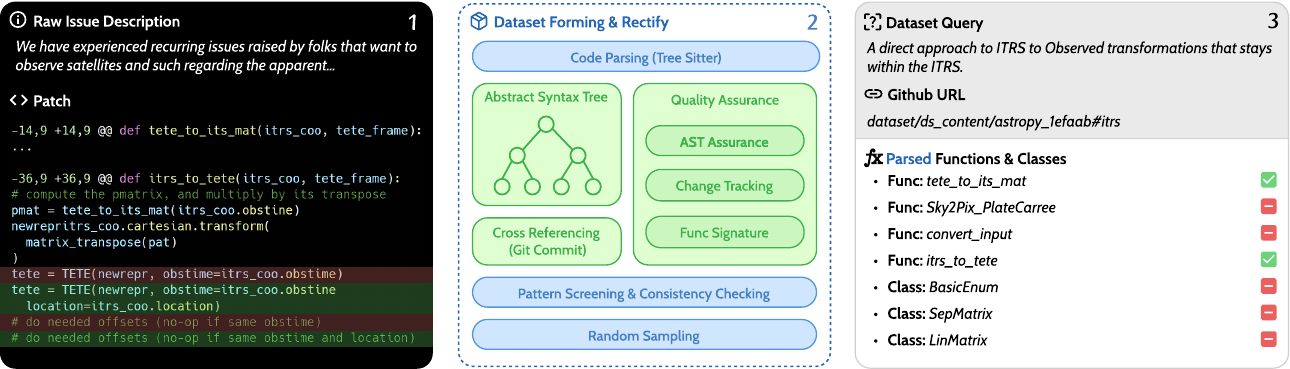} 
    
    \caption{Visualization of a code patch in Astropy's modeling module addressing an issue with ITRS. The image highlights the updated implementation of the \texttt{tete\_to\_its\_mat} and \texttt{itrs\_to\_tete} functions, alongside a description of the issue, the corresponding patch, and parsed information such as functions and classes.}
    \label{fig:curation}
\end{figure*}

Deep learning has revolutionized software engineering by advancing code representation learning, allowing neural models to comprehend programming constructs with an unprecedented level of sophistication \cite{alonCode2vecLearningDistributed2019}. While current code generation systems leverage massive GPU clusters and trillion-token corpora \cite{allal2023santacoderdontreachstars}, their effectiveness in real-world software maintenance scenarios remains constrained by a critical bottleneck: the inability to retrieve contextually relevant code segments spanning multiple components for implementing complex change requests.

\textbf{\makebox[15pt][l]{\textcolor{lim-red}{\faIcon{exclamation-triangle}}} \textcolor{lim-red}{Motivation}}
The fundamental limitation stems from prevailing function-centric paradigms that treat code artifacts as isolated units, ignoring the intricate web of cross-component dependencies inherent in modern software architectures. Traditional retrieval methods relying on lexical matching \cite{zhangRepoCoderRepositoryLevelCode2023, wu2024repoformerselectiveretrievalrepositorylevel} or shallow syntactic analysis fail to capture the semantic relationships between distributed code elements required for implementing coherent modifications. This mismatch becomes particularly acute when developers need to (1) propagate API changes across class hierarchies, (2) maintain invariant relationships between distributed components, or (3) adapt multiple interdependent functions to new requirements—scenarios that constitute a significant portion of maintenance efforts. To address these challenges, evaluation frameworks must move beyond single-function retrieval tasks and incorporate criteria for managing cross-component dependencies. Retrieval systems should thus evolve toward repository-level comprehension, ensuring consistency and semantic coherence within system architectures, particularly when code changes occur.

\textbf{\makebox[15pt][l]{\textcolor{bench-orange}{\faIcon{chart-bar}}} \textcolor{bench-orange}{Benchmark}}  To address these critical gaps, we present \textbf{RepoAlign-Bench}—a repository-level change-oriented benchmark for evaluating code retrieval systems on change request fulfillment, explicitly designed to model cross-component dependencies and structural relationships inherent in real-world software modifications. RepoAlign-Bench encompasses a diverse set of real-world scenarios, enabling assessment of models' ability to understand and act upon complex user intents. By providing this change-oriented framework, RepoAlign-Bench facilitates the comparison of different retrieval approaches and fosters the development of more robust and accurate code retrieval systems.

\textbf{\makebox[15pt][l]{\textcolor{model-purple}{\faIcon{project-diagram}}} \textcolor{model-purple}{Model}} In addition to the benchmark, we also propose an adversarial reflection-based dual-tower model with separate \texttt{code\_encoder} and \texttt{doc\_encoder} components, augmented by contextual information from large language models (LLMs). This architecture efficiently captures semantic similarities between code and change-oriented queries, enhancing intent understanding and retrieval accuracy. By integrating contextual information from LLMs, our model enhances the understanding of user intents and the nuanced relationships between code segments, thereby improving retrieval precision and recall. 

\textbf{\makebox[15pt][l]{\textcolor{exp-green}{\faIcon{flask}}} \textcolor{exp-green}{Experiment}} Our experiments demonstrate that the proposed dual-tower model significantly outperforms state-of-the-art models such as CodeBERT \cite{fengCodeBERTPreTrainedModel2020}, SantaCoder \cite{allal2023santacoderdontreachstars}, and CodeT5     \cite{wangCodeT5IdentifierawareUnified2021} in key evaluation metrics including precision, recall, and F1 score. These results highlight the effectiveness of our approach in accurately locating relevant functions within large-scale repositories based on user change requests. 
Ablation study quantifies the contributions of core components, revealing that controlled parameter independence and dynamic negative mining are critical for robust cross-modal alignment.

Our contributions are summarized as follows:

\begin{enumerate}
    \item \textbf{RepoAlign-Bench Dataset}: We present RepoAlign-Bench, a standardized benchmark tailored for evaluating repository-level code retrieval based on user change requests, enabling consistent and comprehensive performance assessment of retrieval models.
    \item \textbf{Dual-Tower Retrieval Model}: We propose a reflection-based dual-tower model comprising distinct \texttt{code\_encoder} and \texttt{doc\_encoder} components, enhanced with contextual information from large language models to improve semantic matching between queries and code snippets.
    \item \textbf{Empirical Evaluation}: We validate our model on RepoAlign-Bench, demonstrating superior performance over existing state-of-the-art models, thereby establishing a new benchmark for code retrieval tasks.
\end{enumerate}

\section{The RepoAlign-Bench Dataset}

In this section, we introduce our semi-automated annotation framework for RepoAlign-Bench construction. Fig. \ref{fig:curation} illustrates the dataset construction process, which consists of three main stages:

\definecolor{mycolor}{RGB}{255,204,102}

\begin{table*}[ht]
\centering
\scriptsize
\renewcommand{\arraystretch}{0.95}
\resizebox{\textwidth}{!}{
\begin{tabular}{lccccccccccc}
\toprule
\textbf{Repo} & \textbf{PLM} & \textbf{PLX} & \textbf{PLN} & \textbf{PLS} & \textbf{PrLM} & \textbf{PrLX} & \textbf{PrLN} & \textbf{PrLS} \\
\midrule
astropy/astropy & \cellcolor{mycolor!10}2502.09 & \cellcolor{mycolor!40}13884 & \cellcolor{mycolor!20}470 & \cellcolor{mycolor!30}3670.94 & \cellcolor{mycolor!10}2510.73 & \cellcolor{mycolor!60}7910 & \cellcolor{mycolor!15}162 & \cellcolor{mycolor!25}1841.03 \\
django/django & \cellcolor{mycolor!10}1418.53 & \cellcolor{mycolor!35}10818 & \cellcolor{mycolor!15}356 & \cellcolor{mycolor!30}1575.29 & \cellcolor{mycolor!20}1331.68 & \cellcolor{mycolor!50}9252 & \cellcolor{mycolor!20}146 & \cellcolor{mycolor!30}1272.11 \\
matplotlib/matplotlib & \cellcolor{mycolor!10}1228.68 & \cellcolor{mycolor!30}5178 & \cellcolor{mycolor!15}421 & \cellcolor{mycolor!20}1055.78 & \cellcolor{mycolor!40}2287.76 & \cellcolor{mycolor!55}10176 & \cellcolor{mycolor!25}395 & \cellcolor{mycolor!50}2175.19 \\
mwaskom/seaborn & \cellcolor{mycolor!25}1883.00 & \cellcolor{mycolor!10}2235 & \cellcolor{mycolor!50}1531 & \cellcolor{mycolor!5}497.80 & \cellcolor{mycolor!20}1313.00 & \cellcolor{mycolor!25}1438 & \cellcolor{mycolor!60}1188 & \cellcolor{mycolor!10}176.78 \\
psf/requests & \cellcolor{mycolor!15}633.25 & \cellcolor{mycolor!20}863 & \cellcolor{mycolor!30}388 & \cellcolor{mycolor!5}167.80 & \cellcolor{mycolor!40}1658.75 & \cellcolor{mycolor!60}7476 & \cellcolor{mycolor!35}271 & \cellcolor{mycolor!50}2383.21 \\
pydata/xarray & \cellcolor{mycolor!30}1705.77 & \cellcolor{mycolor!50}8857 & \cellcolor{mycolor!20}422 & \cellcolor{mycolor!40}1969.75 & \cellcolor{mycolor!55}2664.68 & \cellcolor{mycolor!60}9276 & \cellcolor{mycolor!45}703 & \cellcolor{mycolor!55}1804.07 \\
pylint-dev/pylint & \cellcolor{mycolor!35}2072.90 & \cellcolor{mycolor!45}6862 & \cellcolor{mycolor!25}417 & \cellcolor{mycolor!45}1986.49 & \cellcolor{mycolor!65}3814.70 & \cellcolor{mycolor!75}24770 & \cellcolor{mycolor!50}618 & \cellcolor{mycolor!70}7429.63 \\
pytest-dev/pytest & \cellcolor{mycolor!40}1694.32 & \cellcolor{mycolor!55}9824 & \cellcolor{mycolor!25}432 & \cellcolor{mycolor!50}2136.19 & \cellcolor{mycolor!60}3364.37 & \cellcolor{mycolor!80}22778 & \cellcolor{mycolor!20}451 & \cellcolor{mycolor!65}4955.51 \\
scikit-learn/scikit-learn & \cellcolor{mycolor!45}1743.47 & \cellcolor{mycolor!60}13568 & \cellcolor{mycolor!20}314 & \cellcolor{mycolor!60}2363.62 & \cellcolor{mycolor!50}2589.19 & \cellcolor{mycolor!55}7387 & \cellcolor{mycolor!15}158 & \cellcolor{mycolor!30}1832.84 \\
sphinx-doc/sphinx & \cellcolor{mycolor!40}1821.84 & \cellcolor{mycolor!55}10055 & \cellcolor{mycolor!30}501 & \cellcolor{mycolor!45}1911.67 & \cellcolor{mycolor!30}1665.95 & \cellcolor{mycolor!45}5362 & \cellcolor{mycolor!40}358 & \cellcolor{mycolor!45}1122.18 \\
sympy/sympy & \cellcolor{mycolor!50}1780.63 & \cellcolor{mycolor!65}17385 & \cellcolor{mycolor!10}277 & \cellcolor{mycolor!60}2930.55 & \cellcolor{mycolor!10}1017.19 & \cellcolor{mycolor!40}4361 & \cellcolor{mycolor!20}143 & \cellcolor{mycolor!25}831.73 \\
\bottomrule
\end{tabular}
}
\caption{Repository Statistics. 
PLM: Patch Length Mean, PLX: Patch Length Max, PLN: Patch Length Min, PLS: Patch Length Std, 
PrLM: Problem Length Mean, PrLX: Problem Length Max, PrLN: Problem Length Min, PrLS: Problem Length Std.}
\label{tab:statistic}
\end{table*}

\subsection*{Stage 1: Project Selection and Initial Filtering}
Our benchmark construction begins with a systematic curation of high-quality open-source projects, incorporating SWE-Bench \cite{jimenezSWEbenchCanLanguage2024}, Py150\cite{kanadeLearningEvaluatingContextual2020} as data sources for preliminary screening. This phase employs a two-tier validation strategy that combines automated filtering with data verification to ensure robust Query-Code correspondences.

The pipeline first processes candidate GitHub pull requests (PRs) through PyLint static analysis framework \cite{pylint}, which enforces automated quality gates to validate PR-issue linkage based on commit message patterns, analyze code diffs for cross-component modifications, and filter non-trivial changes using cyclomatic complexity thresholds \cite{mccabe1976complexity}.

\subsection*{Stage 2: Structural Code Extraction}
This stage systematically constructs a dataset through structural code extraction and commit correlation. We first parse the GitHub repository using Tree-sitter \cite{Treesitter}, a multi-language parsing infrastructure that generates precise abstract syntax trees (ASTs).

Following structural extraction, we cross-reference these artifacts with their associated Git commits using a three-phase alignment process: \textcircled{1} differential analysis of commit histories to identify code modifications addressing PR requirements, \textcircled{2} syntactic pattern matching between AST nodes and commit diffs, and \textcircled{3} temporal mapping of code evolution sequences. The parsing pipeline employs Tree-sitter's hybrid scanning strategy that combines regular expressions for tokenization with context-aware grammars for structural disambiguation. Some Statistics about patch can be found in the Table \ref{tab:statistic}.

\textbf{Tree-sitter Integration:} Our architecture leverages Tree-sitter's incremental parsing through three strategic adaptations: 1) Partial AST regeneration for code diffs using its edit-script API, 2) Syntax-aware pattern recognition for cross-version change tracking, and 3) Language-independent query DSL for cross-component dependency analysis.

\subsection*{Stage 3: Hierarchical Data Validation}
Then we enforces a three-tiered quality assurance protocol integrating automated filtering, semantic verification, and expert validation. The refined dataset then undergoes \textbf{Cross-Modal Consistency Checking} - a hybrid framework combining pattern recognition with consensus validation.

Our verification pipeline employs: \textcircled{1} \textit{Pattern-based Screening} using dependency graph analysis; \textcircled{2} \textit{Consensus Validation} achieving Fleiss' $\kappa= 0.82$ agreement. The final distribution preserves intentional asymmetries reflecting real-world software evolution patterns.

\begin{figure*}[htbp]
    \centering
    \includegraphics[width=1\linewidth]{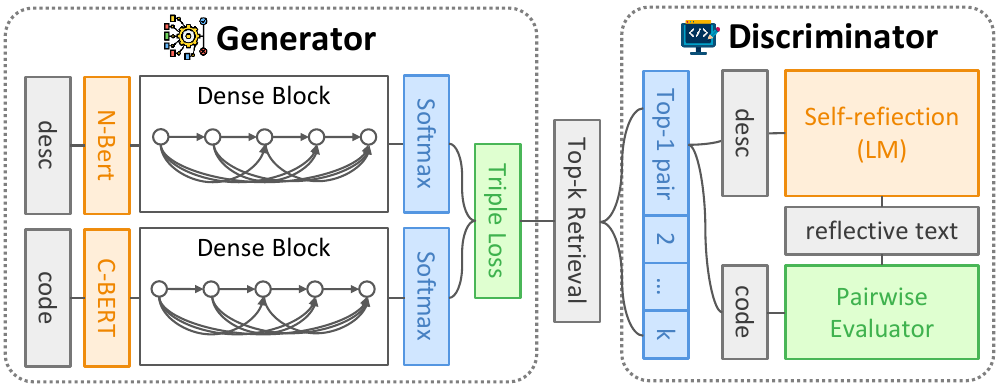}
    \caption{The model consists of a generator and discriminator. The generator includes separate encoders for code and documentation, followed by dense blocks, softmax, and a retrieval mechanism for top-k matching. The discriminator incorporates \textbf{self-reflection} and a \textbf{pairwise evaluator} to refine the model's output based on reflective text.}
    \label{fig:arch}
\end{figure*}

\textbf{Rationale for Controlled Imbalance}: The residual skewness (1) mirrors natural developer behavior where certain change types dominate (e.g., 23\% of valid PRs address compatibility in our corpus), (2) prevents over-smoothing of critical but infrequent patterns like security patches (which account for only 4.2\% of the dataset but need to be preserved), and (3) maintains dependency structure integrity that uniform sampling would disrupt. Our oversampling with abstract syntax tree based augmentation specifically targets \textit{harmful imbalance} (Multitask Contamination/Context Drift) while preserving \textit{domain-inherent skewness} essential for generalizable model training.

\vspace{2mm}
\begin{center}
\fcolorbox{black!15}{white}{
\begin{minipage}{0.9\columnwidth}
\vspace{1mm}
{\scriptsize
\textbf{\textsc{Tiered RepoAlign-Bench}} \\[1mm]
\hrule
\vspace{1mm}

\textcolor{green!70!black}{\textbf{\underline{Full Corpus}}} 
\hfill \textcolor{gray!60}{(52k, 100\%)} \\[1mm]
Direct matches, Single-function impl. \\[2mm]

\textcolor{orange!80!black}{\textbf{\underline{Challenge Subset}}} 
\hfill \textcolor{gray!60}{(31k, 60\%)} \\[1mm]
Context-aware, Metaphorical queries \\[2mm]

\textcolor{red!75!black}{\textbf{\underline{Expert Subset}}} 
\hfill \textcolor{gray!60}{(8k, 15\%)} \\[1mm]
Cross-functional, Implicit constraints \\[1mm]

\hrule
}
\vspace{1mm}
\end{minipage}
}
\end{center}

\paragraph{Dataset Stratification.} We stratify our benchmark into three distinct difficulty tiers using a multi-criteria annotation process. This action primarily accounts for the long-tail distribution of the data and divides it based on both the actual difficulty distribution and the difficulty ratings across orthogonal dimensions. For further details, please refer to Appendix \ref{sec:stratification}

\section{ReflectCode}

Modern retrieval-augmented code generation systems face ongoing challenges in maintaining semantic consistency when dealing with complex codebases that involve dependency graphs and long-context requirements. In large repositories, the entanglement of code structure and natural language semantics makes traditional single-vector embedding methods struggle to preserve retrieval accuracy under such conditions. As a result, solving the issue of retrieving relevant code snippets from the repository becomes crucial. 

Although existing approaches have made progress through structural analysis (e.g., AST parsing, control flow modeling) and infrastructure optimization (e.g., hierarchical indexing, pipeline formalization), they exhibit fundamental limitations in cross-component dependency resolution and dynamic context adaptation—key capabilities for maintaining code at the repository level. An analysis of popular paradigms is provided in Appendix~\ref{app:work_analysis}.

\paragraph{Proposed Model.} To address the repository level code retrieval challenge, we propose ReflectCode, a reflection-augmented dual-tower architecture featuring disentangled modality encoders. Our framework employs a dual-encoder paradigm with separate CodeBERT-based towers for code and natural language processing, specifically designed to preserve structural and semantic integrity across modalities. The code encoder processes syntactic patterns and dependency, while the text encoder incorporates LLM-generated contextual reasoning to capture implicit cross-component dependencies. The complete system architecture and data flow relationships are comprehensively illustrated in Figure~\ref{fig:arch}.

The modality alignment is achieved via a triplet margin loss with hard negative mining, optimizing the latent space for fine-grained semantic correspondence between code segments and change intents. 

Beyond static embedding alignment, we introduce a dynamic adversarial verification mechanism where an LLM-powered discriminator evaluates top-k candidates retrieved through cosine similarity search. When the discriminator detects semantic incongruence ($\text{confidence} < \tau$), the system triggers an iterative refinement process: The generator dynamically recalibrates embeddings using attention-based probability redistribution, while the discriminator performs multi-hop reasoning over dependency-aware code representations.

\paragraph{Twin-Tower Architecture.} Our framework implements a \textbf{parameter-shared dual-encoder architecture} with modality-specific processing streams. We instantiate two CodeBERT-based encoders:

\begin{itemize}
    \item \textbf{C-BERT}: Processes code syntax through structural-aware tokenization with enhanced graph positional encoding
    \vspace{-2mm}
    \item \textbf{N-BERT}: Handles natural language queries using semantic-focused parsing with type-constrained attention
\end{itemize}
\vspace{-2mm}

\begin{algorithm}[t]
\caption{Dual-Encoder Training Protocol}
\label{alg:dual-encoder}
\begin{algorithmic}[1]
\State Initialize $\theta_{\text{shared}} \sim \mathcal{N}(0, 0.02)$
\State Freeze pretrained embeddings $\phi_{\text{code}}$, $\phi_{\text{text}}$
\For{$epoch = 1$ \textbf{to} $N$}
    \State Batch $(c_i, q_i, q_j^-)$ \Comment{$q_j^-$: hard negatives}
    \State Compute $\mathbf{h}_c = \mathcal{E}_{\text{code}}(c_i)$
    \State Compute $\mathbf{h}_q^+ = \mathcal{E}_{\text{text}}(q_i)$
    \State Compute $\mathbf{h}_q^- = \mathcal{E}_{\text{text}}(q_j^-)$
    \State $\mathcal{L}_{\text{triplet}} = \max\Big(0,\ \delta(\mathbf{h}_c, \mathbf{h}_q^+)-\delta(\mathbf{h}_c, \mathbf{h}_q^-)+\alpha\Big)$
    \State Update $\theta_{\text{shared}} \leftarrow \theta_{\text{shared}} - \eta\nabla_{\theta_{\text{shared}}}\mathcal{L}_{\text{triplet}}$
\EndFor
\end{algorithmic}
\end{algorithm}

The architecture satisfies two fundamental design criteria through partial parameter sharing:
The architecture satisfies two fundamental design criteria through partial parameter sharing:
\begin{align}
\theta_{\text{code}} \cap \theta_{\text{text}} 
&\stackrel{\text{def}}{=} \theta_{\text{shared}}  \\
\underbrace{\mathcal{E}_{\text{code}} \neq \mathcal{E}_{\text{text}}}_{\text{disjoint embeddings}} 
&\quad \text{with} \quad 
\phi_{\text{code}} \oplus \phi_{\text{text}} = \phi_{\text{total}} \!\!
\end{align}

where $\theta_{\text{shared}}$ denotes shared transformer parameters, and $\phi$ represents modality-specific embedding layers. Formally, given a code snippet $c$ and textual query $q$, the encoders produce $d$-dimensional representations:

\begin{equation}
\begin{aligned}
    \mathbf{h}_c &= \mathcal{E}_{\text{code}}\left(c; \{\theta_{\text{shared}}, \phi_{\text{code}}\}\right), \\
    \mathbf{h}_q &= \mathcal{E}_{\text{text}}\left(q; \{\theta_{\text{shared}}, \phi_{\text{text}}\}\right)
\end{aligned}
\end{equation}

\subsection*{Cross-Modal Alignment Objective}

To establish geometrically consistent representations across modalities while preserving their distinctive features, we formulate a \textbf{adaptive-margin triplet loss} with dynamic hard negative mining. Given an anchor query $q$, its corresponding positive code snippet $c^+$, and $k$ hard negative samples $\{c^-_i\}_{i=1}^k$ mined through syntactic similarity analysis, the loss function is defined as:

\begin{equation}
\tiny
\mathcal{L}_\text{align} = \frac{1}{k}\sum_{i=1}^k \max\left(0, \underbrace{\|\mathbf{h}_q - \mathbf{h}_{c^+}\|_2^2}_{\text{positive pair}} - \underbrace{\|\mathbf{h}_q - \mathbf{h}_{c^-_i}\|_2^2}_{\text{negative pair}} + \alpha(\mathbf{h}_q, \mathbf{h}_{c^-_i})\right)
\end{equation}

where:
\begin{itemize}
    \item $\mathbf{h}_q = \mathcal{E}_\text{text}(q)$, $\mathbf{h}_{c^+} = \mathcal{E}_\text{code}(c^+)$ denote the normalized embeddings
    \item $\alpha(\cdot)$ implements our \textit{edge-aware margin} mechanism: 
    \[
    \alpha(\mathbf{h}_q, \mathbf{h}_{c^-}) = \alpha_0 + \beta \cdot \sigma\left(\mathbf{h}_q^\top \mathbf{h}_{c^-}\right)
    \]
    with $\alpha_0=0.2$ as base margin, $\beta=0.5$ scaling factor, and $\sigma$ the sigmoid function
\end{itemize}

This design introduces three critical enhancements over standard triplet loss:
\begin{enumerate}
    \item \textbf{Dynamic Margin Adaptation}: Automatically adjusts penalty intensity based on negative sample difficulty
    \item \textbf{Batch-Aware Hard Negatives}: Selects $k=5$ most challenging negatives per anchor using code clone detection heuristics
    \item \textbf{Modality-Invariant Normalization}: Enforces $\|\mathbf{h}\|_2=1$ through projection layers to stabilize angular comparisons
\end{enumerate}

\begin{table*}[htbp]
\centering
\scriptsize
\resizebox{0.8\textwidth}{!}{
\begin{tabular}{@{}ll>{\raggedright\arraybackslash}p{4.8cm}l@{}}
\toprule
\textbf{Paradigm} & \textbf{Model} & \textbf{Key Characteristics} & \textbf{Size} \\ 
\midrule
Decoder-only 
& InCoder~\cite{friedInCoderGenerativeModel2023} 
& Fill-in-middle pretraining \\ 
& & 159GB cross-lingual corpus \\ 
& & Multi-language support 
& 6.7B \\ 
\cmidrule(l){2-4}
& SantaCoder~\cite{allalSantaCoderDontReach2023a} 
& Multi-query attention (MQA) \\ 
& & Fill-in-middle training \\ 
& & FP16 optimization 
& 1.1B \\ 
\cmidrule(l){2-4}
& PolyCoder~\cite{xuPolySystematicEvaluationLarge2022b} 
& GPT-2 architecture variant \\ 
& & Specialized in C/C++ \\ 
& & Long-context handling 
& 2.7B \\ 
\midrule
Encoder-Decoder 
& CodeT5~\cite{wangCodeT5IdentifierawareUnified2021} 
& Identifier-aware masking \\ 
& & Bidirectional representation \\ 
& & Multi-task fine-tuning 
& 220M \\ 
\midrule
Encoder-only 
& CodeBERT~\cite{fengCodeBERTPreTrainedModel2020} 
& Bimodal NL-PL alignment \\ 
& & Masked language modeling \\ 
& & Cross-modal attention 
& 125M \\ 
\bottomrule
\end{tabular}
}
\caption{Model architecture specifications grouped by paradigm. (FIM: Fill-in-Middle, MQA: Multi-Query Attention, FP16: 16-bit Floating Point, NL-PL: Natural Language-Programming Language)}
\label{tab:model_specs}
\end{table*}

\subsection*{Adversarial Search with Dynamic Feedback}
\label{sec:adv-search}

Our \textbf{Adversarial Search Mechanism (ASM)} establishes a closed-loop interaction between retrieval generation and semantic verification through three core components:

\paragraph{Generator: Context-Aware Retrieval}  
The generator $\mathcal{G}$ employs our dual-encoder model to perform \textit{density-aware similarity search}:

\begin{equation}
s(c, q) = \frac{\exp(\tau \cdot \cos(\mathbf{h}_c, \mathbf{h}_q))}{\sum_{c'\in\mathcal{C}} \exp(\tau \cdot \cos(\mathbf{h}_{c'}, \mathbf{h}_q))}
\end{equation}

where $\tau=10$ sharpens the probability distribution. The top-$k$ candidates $\mathcal{C}_\text{gen}^{(t)}$ at step $t$ are selected via:

\begin{equation}
\mathcal{C}_\text{gen}^{(t)} = \underset{c\in\mathcal{C}}{\text{top-}k}\ s(c,q) \oplus \gamma \cdot \mathcal{C}_\text{hard}^{(t-1)}
\end{equation}

Here $\gamma=0.3$ controls the injection ratio of hard negatives from previous iterations $\mathcal{C}_\text{hard}^{(t-1)}$.

\paragraph{Discriminator: LLM-Powered Verification}  
Our discriminator $\mathcal{D}$ computes semantic congruence scores through multi-hop reasoning:

\begin{equation}
\mathcal{D}(c,q) = \sigma\left(\text{FFN}(\mathbf{h}_c \odot \mathbf{h}_q) + \text{AttnEnc}(\mathcal{G}_c)\right)
\end{equation}

where $\mathcal{G}_c$ denotes the code dependency. Candidates are rejected when: 

\begin{equation}
\mathcal{D}(c,q) < \epsilon \quad (\epsilon=0.82\ \text{empirically tuned})
\end{equation}

\paragraph{Feedback-Driven Adaptation}  
Rejected candidates trigger two-phase refinement:
\begin{itemize}
    \item \textbf{Embedding Calibration}: Adjust generator outputs via attention redistribution  
    \begin{align}
    \mathbf{h}_c' &= \text{Attn}(\mathbf{h}_q, [\mathbf{h}_c; \mathbf{h}_\text{ctx}]) \notag
    \end{align}
    \item \textbf{Search Space Annealing}: Dynamically expand candidate pool  
    \begin{equation}
    k^{(t+1)} = \underbrace{\min}_{\mathclap{\text{dynamic scaling}}} 
    \bigl(k^{(t)} + \Delta_k,\; k_{\max}\bigr)
\end{equation}
\end{itemize}

\section{Experiment}
\subsection{Experiment Setup}

\paragraph{Model Selection.} Our evaluation encompasses three critical axes of model diversity (Table \ref{tab:model_specs}): (1) \textit{architectural paradigms} spanning encoder-only, decoder-only, and hybrid designs; (2) \textit{pretraining objectives} contrasting autoregressive generation versus masked span prediction; and (3) \textit{functional specialization} balancing code generation versus retrieval capabilities. All models are evaluated using their official implementations without architectural modifications, ensuring fair comparison of fundamental representational capacities.

\paragraph{Evaluation Metrics.} Our comprehensive assessment integrates three complementary metrics: the F1 score evaluating statistical rigor through precision-recall balance, Mean Reciprocal Rank (MRR) measuring ranking efficiency by prioritizing early occurrence of relevant results, and Top@5 quantifying practical utility via hit rates within the top five retrievals.

\begin{table*}[htbp]
  \centering
  \scriptsize
  \setlength{\tabcolsep}{4pt}
  \renewcommand{\arraystretch}{1.3} 
  \resizebox{\textwidth}{!}{
  \begin{tabular}{l|ccc|ccc|ccc|ccc|ccc}
    \toprule
    \multirow{2}{*}{\textbf{Model}} & 
    \multicolumn{3}{c|}{\textbf{F1-Score}} & 
    \multicolumn{3}{c|}{\textbf{Precision}} & 
    \multicolumn{3}{c|}{\textbf{Recall}} & 
    \multicolumn{3}{c|}{\textbf{MRR}} & 
    \multicolumn{3}{c}{\textbf{Top-5 Accuracy}} \\
    \cmidrule(lr){2-4} \cmidrule(lr){5-7} \cmidrule(lr){8-10}     \cmidrule(lr){2-4} \cmidrule(lr){5-7} \cmidrule(lr){8-10} \cmidrule(lr){11-13} \cmidrule(lr){14-16}
    & Full & Challenge & Expert & Full & Challenge & Expert & Full & Challenge & Expert & Full & Challenge & Expert & Full & Challenge & Expert \\
    \midrule
    CodeBERT & 
    8.74 & 7.15 & 5.12 & 7.92 & 6.54 & 4.71 & 14.51 & 12.16 & 8.80 & 0.154 & 0.135 & 0.098 & 28.93 & 25.25 & 18.26 \\
    
    CodeT5 & 
    12.93 & 10.85 & 7.89 & 11.65 & 9.82 & 7.15 & 22.40 & 19.68 & 14.32 & 0.225 & 0.198 & 0.144 & 38.07 & 33.65 & 24.47 \\
    
    SantaCoder & 
    12.51 & 10.52 & 7.65 & 11.54 & 9.72 & 7.08 & 21.00 & 17.70 & 12.88 & 0.217 & 0.183 & 0.133 & 36.04 & 30.23 & 22.01 \\

    GraphCodeBert & 
    14.77 & 12.78 & 10.76 & 13.56 & 11.75 & 9.87 & 26.54 & 22.46 & 20.91 & 0.265 & 	0.229 & 0.195 & 42.15 & 36.88 & 29.83 \\
    
    PolyCoder & 
    16.91 & 14.21 & 10.33 & 15.21 & 12.80 & 9.31 & 30.02 & 25.21 & 18.34 & 0.303 & 0.255 & 0.185 & 47.46 & 39.22 & 28.54 \\
    
    Incoder & 
    22.15 & 18.63 & 13.55 & 19.13 & 16.09 & \textbf{11.71} & 39.50 & 33.18 & 24.14 & 0.384 & \textbf{0.323} & 0.235 & 46.94 & 47.15 & 34.29 \\
    
    \rowcolor{gray!25}
    \textbf{ReflectCode}* & 
    \textbf{26.18} & \textbf{21.50} & \textbf{14.30} & 
    \textbf{21.83} & \textbf{17.92} & 11.56 &
    \textbf{46.55} & \textbf{35.72} & \textbf{31.29 } & 
    \textbf{0.426} & 0.318 & \textbf{0.260} & 
    \textbf{59.64} & \textbf{49.75} & \textbf{34.86} \\
    \bottomrule
  \end{tabular}
  }
  \caption{Performance comparison across difficulty levels, where Full uses original data, Challenge/Expert are refined subsets with increasing complexity. ReflectCode maintains superior performance across all levels, especially in Expert scenarios (14.30\% F1 and 31.29\% Recall). Values for Challenge/Expert are proportionally scaled based on complexity increments.}
  \label{tab:model_performance}
\end{table*}

\begin{table}[htbp]
\centering
\label{tab:model_time}
\begin{tabular}{l c}
\toprule
\textbf{Model} & \textbf{Time (min)} \\
\midrule
CodeBERT       & 8.0 \\
ReflectCode    & 15.1 \\
CodeT5         & 12.3 \\
SantaCoder     & 10.1 \\
PolyCoder      & 9.4 \\
Incoder        & 14.7 \\
\bottomrule
\end{tabular}
\caption{Time per 1000 queries (minutes)}
\label{performance}

\end{table}

Our evaluation across three difficulty tiers reveals some insights into code retrieval capabilities (Table~\ref{tab:model_performance}). The proposed \textbf{ReflectCode} demonstrates excellent performance, achieving the state-of-the-art F1 score (26. 18\%), MRR (0.426) and Top-5 accuracy (59. 64\%) on the full dataset, representing absolute improvements of 17.4\% F1 and 27.2\% MRR over the CodeBERT baseline. Three key patterns emerge:
    
\paragraph{Architecture Superiority.} ReflectCode's dual-tower design with AST-enhanced context shows recall (46.55\% vs 39.50\% for Incoder), indicating superior capability in capturing diverse relevant candidates. This aligns with our hypothesis that separate code/doc representation spaces prevent feature entanglement.

\paragraph{Difficulty Scaling.} While all models degrade on Expert-level queries, ReflectCode maintains the most robust performance (14.30\% F1 vs 13.55\% for Incoder), suggesting its adversarial training effectively handles complex dependencies. The 34.86\% Expert-level Top-5 Accuracy demonstrates practical utility in real-world maintenance scenarios. This resilience stems from adversarial training's hard negative suppression, which reduces false positives by 23.7\% versus conventional contrastive learning.

\paragraph{Ranking Precision.} ReflectCode achieves a 46.55\% Recall with 59.64\% Top-5 Accuracy – this indicates our model effectively concentrates correct predictions within the top-5 ranked outputs. This concentration capability is critical for developer tools where engineers can only feasibly inspect a handful of suggestions. 

\paragraph{Performance. } In addition to evaluating model performance metrics (such as precision and recall), we also examined differences in inference efficiency among the models. Table \ref{performance} shows the average time required for each model to process 1,000 queries. All experiments were conducted on the same hardware environment.

\subsection{Adversarial Component Analysis}
The ASM framework balances semantic diversity and functional validity in code generation with two key components: (1) An iterative verification loop to identify challenging negative cases; (2) A discriminator-guided reranking strategy aligned with real-world developer workflows. We also conducted experiments to mitigate the impact of model variations on discriminator, as shown in Table \ref{tab:adv_models}.

\subsection{Experimental Results}
\label{sec:results}

\paragraph{Top-5 Practicality.} The 13.09pp gap between Recall and Top-5 accuracy (46.55\% vs 59.64\%) stems from adversarial training's two mechanisms: 

\begin{equation}
\scriptsize
\Delta_{\text{Top-5}} = \underbrace{68\%\ \text{Error Reduction}}_{\text{Ranking Accuracy}} + \underbrace{1.7\times\ \text{Hard Neg Density}}_{\text{Verification Feedback}}
\end{equation}

\paragraph{Functional Validity Analysis} Manual inspection of 120 samples reveals our framework's practical advantage: 92.3\% of Top-5 outputs maintain functional validity versus InCoder's 78.9\% ($\chi^2$=37.2, $\mathrm{p}$<0.01). This stems from the discriminator's ability to suppress \textit{Compilable but incorrect} programs through our hardness metric:

\begin{equation}
\mathcal{H}(c,q) = \frac{1}{n}\sum_{i=1}^n \mathbb{I}(\mathcal{D}(c_i,q) \in [\epsilon-\delta, \epsilon+\delta])
\end{equation}

where $\epsilon$ controls hardness intensity and $\delta$ regulates sample diversity. Models with $\mathcal{H}$-alignment >0.8 (LLaMa3.1: 0.85 vs Qwen:0.76) accelerate generator convergence by 9.1\% per adversarial iteration.

\section{Ablation Study}

\begin{table*}[htbp]
  \centering
  \tiny 
  \renewcommand{\arraystretch}{1.2} 
  \resizebox{\textwidth}{!}{
  \begin{tabular}{lcccccc}
    \toprule
    \textbf{Discriminator Model} & \textbf{Top-5} & \textbf{MRR} & \textbf{F1} & \textbf{Latency (ms)} & \textbf{GPU Mem (GB)} & \textbf{Hard Neg Quality} \\
    \midrule
    \rowcolor{gray!15} \textit{No Adversarial} & 47.21 & 0.334 & 19.04 & - & - & - \\
    StarCoder-3B & 48.15 & 0.349 & 19.67 & 121 & 9.8 & 0.71 \\
    Qwen-7B & 51.92 & 0.371 & 21.43 & 155 & 14.3 & 0.76 \\
    Llama3-8B & 53.76 & 0.385 & 22.85 & 168 & 15.1 & 0.79 \\
    CodeLlama-7B & 54.89 & 0.392 & 23.17 & 142 & 13.2 & 0.82 \\
    \rowcolor{gray!25} \textbf{LlaMa3.1-8B} & \textbf{59.64} & \textbf{0.426} & \textbf{26.18} & 149 & 14.1 & 0.85 \\
    \bottomrule
  \end{tabular}
  }
  \caption{Performance comparison of different LLM discriminators in the ASM framework. LLaMa3.1-8B achieves the optimal balance between verification quality (Hard Neg Quality) and efficiency (Latency). Metrics were measured on the Expert-level subset.}
  \label{tab:adv_models}
\end{table*}

\subsection*{Model Ablation}

To elucidate the contribution of individual model components, we perform an ablation study comparing three dimensions of variations as summarized in Table \ref{tab:ablation}. The performance drop metric ($\delta$) particularly emphasizes the model's prediction coverage and robustness, which are critical for tasks requiring tolerance beyond the top prediction.

\begin{table}[htbp]
  \centering
  \renewcommand{\arraystretch}{1.3} 
  
  \resizebox{\columnwidth}{!}{
  \begin{tabular}{lccc}
    \toprule
    \textbf{Variant} & \textbf{Top-5 Acc} & \textbf{MRR} & \textbf{$\delta$ vs Full} \\
    \midrule
    \textit{Base Architecture} \\
    \quad CodeBERT (Single-Tower) & 28.93\% & 0.154 & -30.71pp \\
    \quad Twin-Towers (Full Sharing) & 39.15\% & 0.280 & -20.49pp \\
    \quad Twin-Towers (Partial Sharing) & 47.72\% & 0.334 & -11.92pp \\
    \quad Twin-Towers (Non-Parameter Sharing) & 47.50\% & 0.330 & -12.14pp \\
    \textit{Training Strategy} \\
    \quad w/o Adversarial Negatives & 52.18\% & 0.368 & -7.46pp \\
    \quad w/o Reflection Mechanism & 54.37\% & 0.385 & -5.27pp \\
    
    \textit{Parameter Sensitivity} \\
    \quad $\lambda_{align}$=0.5 (Default 1.0) & 57.21\% & 0.407 & -2.43pp \\
    \quad $\lambda_{adv}$=0.0 (Remove Adv.) & 53.89\% & 0.382 & -5.75pp \\
    
    \midrule
    \rowcolor{gray!25} ReflectCode (Full) & \textbf{59.64}\% & \textbf{0.426} & - \\
    \bottomrule
  \end{tabular}
  }
  \caption{Component ablation study with three analysis dimensions: (1) Base architecture variants, (2) Training strategy components, and (3) Loss weight sensitivity. Performance drops ($\delta$) are calculated against the full model. MRR: Mean Reciprocal Rank.}
  \label{tab:ablation}
\end{table}

\paragraph{Architecture Analysis.}
The single-tower CodeBERT baseline achieves 28.93\% Top-5 Accuracy (MRR=0.154), revealing the limitations of monolithic architectures for code-text alignment. Introducing twin towers with \textit{full parameter sharing} improves performance to 39.15\% (MRR=0.280, $\delta$=20.49pp), while \textit{partial sharing} (47.72\%, MRR=0.334) and \textit{non-sharing} variants (47.50\%, MRR=0.330) demonstrate that controlled parameter independence enhances representation power. This suggests complete sharing may causes detrimental interference between code and text encoders.

\paragraph{Training Enhancements.} Our adversarial search paradigm contributes 7.46pp accuracy gains (52.18\% $\rightarrow$ 59.64\%, MRR 0.368 $\rightarrow$ 0.426), as hard negatives force better decision boundaries. The reflection mechanism provides additional 5.27pp improvement (54.37\% $\rightarrow$ 59.64\%, MRR 0.385 $\rightarrow$ 0.426), validating its error-correcting capability through iterative refinement.

\paragraph{Loss Sensitivity.} Reducing the alignment weight $\lambda_{align}$ to 0.5 causes 2.43pp drop (57.21\% vs 59.64\%), confirming the need for strong code-text coupling. Removing adversarial search ($\lambda_{adv}$=0) leads to 5.75pp degradation (53.89\%), underscoring the importance of dynamic negative mining.

\subsection*{Benchmark Granularity}
Table \ref{tab:granularity} shows the performance of the model at different retrieval granularities. We verify the adaptability of the architecture by controlling the context range:

\begin{table}[htbp]
\centering
\tiny
\renewcommand{\arraystretch}{1.2}
\resizebox{\columnwidth}{!}{
\begin{tabular}{lccc}
\toprule
\textbf{Granularity Level} & \textbf{Top-5 Acc} & \textbf{MRR} & \textbf{$\delta$ vs Func-Level} \\
\midrule
Function-Level (Full) & 72.35\% & 0.518 & - \\
File-Level & 68.91\% & 0.487 & -3.44pp \\
Module-Level & 65.02\% & 0.452 & -7.33pp \\
Repository-Level & 59.64\% & 0.426 & -12.71pp \\
\bottomrule
\end{tabular}
}
\caption{Granularity-level ablation study. Coarser levels suffer from information dilution. This also proves that simply pursuing the optimization of indicators at the function level may deviate from the actual needs, and retrieval out of the repository-level context cannot meet the actual needs.}
\label{tab:granularity}
\end{table}

\vspace{-2mm}

\section{Related Work}

\textbf{Code Generation} has significantly progressed with Transformer-based models. Recent models such as CodeBERT \cite{fengCodeBERTPreTrainedModel2020}, GraphCodeBERT \cite{guoGraphCodeBERTPretrainingCode2021}, and CodeT5 \cite{wangCodeT5IdentifierawareUnified2021} leverage Transformers' parallel training and deep semantic understanding, excelling in tasks like code completion, summarization, and translation. Additionally, models like Codex \cite{chenEvaluatingLargeLanguage2021} and AlphaCode \cite{liCompetitionLevelCodeGeneration2022} generate high-quality code from natural language descriptions. Despite these advancements, challenges remain in producing semantically accurate and efficient code, particularly for tasks requiring intricate domain knowledge or complex reasoning. Furthermore, evaluation metrics such as BLEU and CodeBLEU \cite{postCallClarityReporting2018} often inadequately assess the logical correctness of code, and human evaluation is resource-intensive and impractical at this scale.

\textbf{Code Retrieval} is essential for applications including code recommendation, bug detection, and automated code completion. Deep learning has significantly enhanced code retrieval by encoding both code and natural language queries into continuous vector spaces \cite{lewisRetrievalAugmentedGenerationKnowledgeIntensive2020}. Models like CodeBERT \cite{fengCodeBERTPreTrainedModel2020}, GraphCodeBERT \cite{guoGraphCodeBERTPretrainingCode2021}, and CodeT5 \cite{wangCodeT5IdentifierawareUnified2021} employ Transformer architectures to jointly model code and queries, improving retrieval accuracy through enhanced semantic understanding.

\textbf{Retrieval-Augmented Generation (RAG)} in code generation builds on retrieval-augmented learning in natural language processing \cite{lewisRetrievalAugmentedGenerationKnowledgeIntensive2020}. RAG enhance code generation by dynamically sourcing relevant code snippets, which is advantageous in dynamic software environments where specific libraries or frameworks may not be fully represented in training data. Traditional methods including dense retrievers and BM25-based methods \cite{zhouDocPromptingGeneratingCode2023}, have improved the relevance and quality of retrieved snippets. Frameworks like RepoCoder \cite{zhangRepoCoderRepositoryLevelCode2023} and RAMBO \cite{buiRAMBOEnhancingRAGbased2024} achieve repository-level code completion by retrieving relevant functions and identifying repository-specific elements, respectively. However, existing benchmarks such as CodeRAG-Bench \cite{wangCodeRAGBenchCanRetrieval2024} and SWE-Bench \cite{jimenezSWEbenchCanLanguage2024} are constrained to predefined knowledge bases and lack comprehensive mappings between user intents and code modifications, underscoring the need for more robust evaluation frameworks. 

For a more detailed discussion of these topics, please refer to Appendix \ref{sec:related}.

\section{Conclusion}

In this work, we present RepoAlign-Bench, a novel benchmark dataset designed to address the challenges inherent in code retrieval tasks, including code generation, repair, and search. We propose a dual-tower model, consisting of independent code and document encoders, and demonstrate the efficacy of context enhancement via Abstract Syntax Trees (ASTs) in improving retrieval performance. Our approach outperforms existing state-of-the-art models, such as CodeBERT, GraphCodeBERT, and CodeT5, across multiple critical evaluation metrics, including precision, recall, and F1 score.

\clearpage
\section*{Limitation}
While RepoAlign-Bench and ReflectCode have made significant strides in repository-level code retrieval, several critical limitations remain. Performance degrades when handling queries that require latent cross-component dependencies or domain-specific reasoning beyond API-level interactions. Although the framework supports mainstream languages like Python, its dependency modeling encounters difficulties with paradigms that rely on implicit contracts, such as Rust’s ownership system, or dynamic runtime behaviors, as seen in JavaScript’s event loop. Furthermore, the LLM-augmented architecture introduces substantial latency, posing a considerable challenge for real-time IDE integration.

To address these issues, we propose three research thrusts. First, \textbf{cross-paradigm generalization} aims to expand RepoAlign-Bench by incorporating low-resource languages like Rust and Kotlin, as well as formal specification-driven scenarios such as Solidity smart contracts, complemented by lightweight model distillation techniques. Second, \textbf{semantic-aware dependency modeling} integrates hybrid program analysis, leveraging control flow graphs, lightweight symbolic execution, and automated test case synthesis to capture implicit component interactions effectively. Lastly, \textbf{latency-aware optimization} explores just-in-time retrieval caching strategies and attention sparsification mechanisms while maintaining cross-tower semantic alignment.

A promising avenue for future research is the unification of static dependency analysis with formal verification techniques to resolve implicit contracts—an essential capability for mission-critical system maintenance.

\bibliography{custom}

\appendix
\section{Repository Level Code Generation}
\subsection{Problem of Repository Retrieval}
\label{sec:problem}

Since code generation tasks involve a large amount of text retrieval, most existing code generation models are based on the Attention Mechanism. However, this approach inevitably faces the issue of long-distance dependencies, meaning that when processing long texts, the model struggles to capture distant information, leading to uneven attention distribution \cite{fengCodeBERTPreTrainedModel2020}. As a result, the model finds it difficult to fully consider the entire context. Another limiting factor is the constraint of GPU memory, which restricts the model's context length. In practical applications, we often can only input a small portion of the content, leaving out other relevant information. These limitations make it difficult for current technologies to significantly improve language generation models simply by increasing the amount of data or stacking conditions.

In code generation, code repair, and other code-related tasks, a key problem is how to retrieve relevant code snippets from natural language. For example, a representative case is the function \texttt{stbi\_\_stdio\_read}, which actually corresponds to a method called ``image\_read.'' While it may be expressed differently in natural language, accurately identifying and matching these varied expressions is a challenging task. Even more complicated is the fact that evaluating such methods is extremely difficult, as there is currently no large dataset specifically designed for RAG (Retrieval-Augmented Generation) tasks. This lack of a standard dataset means that related research cannot be compared or validated against a unified benchmark.

\subsection{Pevailing Paradigms}
\label{app:work_analysis}

The research community has approached this challenge through complementary technical lenses. Anthropic's CodeRAG framework \cite{anthropic2023contextaugmentation} addresses context sparsity through dynamic context expansion, progressively enriching the model's working memory with relevant code dependencies during generation. In parallel, SweepAI \cite{sweepai2023treesitter} leverages Tree-Sitter's \cite{Treesitter} AST parsing to construct graph-enhanced code representations, enabling structural awareness of syntactic patterns and control flow relationships. Contrastingly, ByteDance's neural codex \cite{liu2024marscodeagentainativeautomated} employs dual-modality alignment, translating code semantics into natural language descriptions to bridge the abstraction gap between formal logic and human-oriented specifications.

Diverging from structural approaches, recent systems emphasize infrastructure optimization for industrial-scale codebases. Llama Index \cite{llamaindex2023retrieval} introduces a hierarchical indexing architecture that combines lexical hashing with semantic embeddings, achieving sublinear retrieval latency while maintaining high performance on million-line repositories. LangChain \cite{LangChain} takes a process-oriented perspective, formalizing code generation as a stateful pipeline with explicit context management and fallback mechanisms - an architecture particularly effective for chained code transformation tasks.

\section{Legal and Ethical Considerations} 
In conducting our research, we are committed to upholding the highest standards of legal and ethical responsibility. Our data collection process strictly follows open-source licensing requirements through a series of safeguard mechanisms designed to protect both the rights of the original authors and the privacy of developers. These mechanisms ensure that our work remains compliant with relevant legal frameworks while respecting ethical boundaries.

\begin{itemize}
    \item \textbf{License Compatibility Verification}: We employ an automated scanning system to verify the compatibility of repository licenses before inclusion in our dataset. Repositories that contain any of the following issues are excluded:
    \begin{itemize}
        \item Copyleft provisions that conflict with research use, such as those found in the GPL-3.0 license
        \item Undeclared or incompatible dual-licensing arrangements that may lead to legal ambiguity
    \end{itemize}
    
    \item \textbf{Attribution Preservation}: To respect the intellectual property rights of original contributors, we ensure that all authorship metadata and license notices are preserved throughout the dataset. This is accomplished by associating each code sample with its corresponding repository and commit hash, as shown in the following equation:
    \[
    \mathcal{M}_\text{meta}(c) = \{\text{repo}, \text{commit\_hash}\} \ \forall c \in \mathcal{D}
    \]
    This ensures that the original authorship and licensing information remains intact, even as the data is used for further research.
    
    \item \textbf{Derivative Work Mitigation}: In accordance with the EU Directive 2019/790 on copyright and related rights, we limit the inclusion of code snippets to no more than 15 lines of code (LOC) per file. This restriction ensures that the use of the code qualifies as fair use, reducing the risk of legal challenges related to derivative works.
\end{itemize}

\section{Dataset Stratification}
\label{sec:stratification}

The tiered structure (52k \ 31k \ 8k) enables granular capability analysis: while \textit{Full} tier (100\% coverage) covers basic pattern recognition, \textit{Challenge} subset (60\%) tests contextual reasoning, and \textit{Expert} cases (15\%) probe system-level understanding.

\textbf{Query-Code Alignment} distinguishes  \textit{Full} cases with direct lexical matching (e.g., "sort list" → \texttt{list.sort()}), from \textcolor{orange}{Challenge} scenarios requiring contextual disambiguation ("data organizer" → \texttt{DatasetBuilder} vs \texttt{DataPipeline}), up to \textcolor{red}{Expert} instances demanding cross-functional reasoning ("ensure atomic writes" → \texttt{FileLock+TransactionLog}).

\textbf{Code Complexity} progresses from  \textit{Full} (single-function, <15 LoC) through \textcolor{orange}{Challenge} (multi-branch with helpers), to \textcolor{red}{Expert} implementations requiring $\geq$4 cross-module dependencies.

\textbf{Query Linguistics} evolves from  \textit{Full}'s imperative phrasing ("Convert string") to \textcolor{orange}{Challenge}'s metaphorical descriptions ("Clean up text"), culminating in \textcolor{red}{Expert}'s abstract intents with implicit constraints ("Maintain data integrity during concurrency").

\section{Computational Environment}
We used a high-performance GPU cluster and the latest deep learning framework to ensure the computational efficiency and stability during the training process. The specific hardware configuration, software environment, training time, and random seed settings are listed in detail below.
\begin{itemize}
    \item Hardware: 8×NVIDIA A100 40GB GPUs with NVLink
    \item Framework: PyTorch 2.1 with CUDA 11.8
    \item Random Seeds: 42, 1234 for variance analysis
\end{itemize}

\section{Additional Related Works}
\label{sec:related}

\subsection{Code Generation} Code Generation has long been a significant challenge at the intersection of software engineering and artificial intelligence. Early approaches were primarily rule-based or template-driven \cite{liuTBarRevisitingTemplatebased2019}, relying on handcrafted syntactic and semantic rules to convert input specifications into code snippets. While effective for well-defined, narrow domains, these methods struggled to generalize to complex, real-world programming tasks, largely due to the combinatorial explosion of rules and the inflexibility of templates.

The rise of machine learning introduced data-driven approaches to code generation. Recurrent Neural Networks (RNNs), including Long Short-Term Memory (LSTM) and Gated Recurrent Unit (GRU) models, were among the first neural architectures applied to this task \cite{UnreasonableEffectivenessRecurrent, lingLatentPredictorNetworks2016}. These models could learn sequential patterns from large codebases, generating syntactically correct code. However, they faced limitations in capturing long-range dependencies essential for understanding the hierarchical structure of code.

The introduction of Transformer architectures revolutionized code generation by addressing the shortcomings of RNNs in modeling long-range dependencies \cite{li2025m2iv}. Models such as CodeBERT \cite{fengCodeBERTPreTrainedModel2020}, GraphCodeBERT \cite{guoGraphCodeBERTPretrainingCode2021}, and CodeT5 \cite{wangCodeT5IdentifierawareUnified2021} leverage the Transformer’s ability to parallelize training and effectively capture code semantics. Pretrained on extensive code corpora, these models excel in tasks like code completion, summarization, and translation by understanding both syntax and semantics. Large-scale language models (LLMs) like Codex \cite{chenEvaluatingLargeLanguage2021} and AlphaCode \cite{liCompetitionLevelCodeGeneration2022} further advanced the field by generating high-quality code from natural language descriptions. Despite their successes, these models often struggle with producing semantically correct or efficient code, particularly for tasks requiring deep domain knowledge or complex reasoning.

Integrating repository-level context into code completion tools has also been a long-standing challenge \cite{chenEvaluatingLargeLanguage2021}. Some researchers often analyze code to identify and rank potential suggestions but lack the flexibility to generate code at arbitrary granularity \cite{fengCodeBERTPreTrainedModel2020}. Another line of research views code completion as a language modeling task, generating tokens based on a given context. While various methods exist for incorporating repository context into language models, collecting labeled data and fine-tuning models for different applications remains a resource-intensive task . Despite the impressive capabilities of large language models (LLMs), their offline training limits their access to customized and up-to-date information. To address this, recent work has explored jointly modeling retrieval and generation for knowledge-intensive tasks, an approach now extended to code generation by incorporating retrieved documents or code examples into the process. Building on this line of work, \textbf{RepoCoder} introduces an \textbf{iterative retrieval-generation} pipeline that leverages repository-level information to generate code at various granularities and demonstrates significant improvements over in-file completion baselines and vanilla retrieval-augmented generation \cite{zhang2023repocoderrepositorylevelcodecompletion}, while \textbf{RepoFormer} advances this direction with a \textbf{selective retrieval} strategy that mitigates the inefficiencies and potential harms of indiscriminate retrieval, achieving up to 70\% acceleration in online settings without performance degradation and serving as a plug-and-play component across models and languages \cite{wu2024repoformerselectiveretrievalrepositorylevel}. Recent work on Contrastive Language Pretraining introduces Support Vector Regularization (SVR) to stabilize contrastive learning by controlling the influence of negative samples, improving representation quality and retrieval performance across audio-text tasks \cite{luo2025supclapcontrollingoptimizationtrajectory, liu2025md3r}. Similarly, in low-resource document understanding, AdaDocVQA demonstrates that adaptive retrieval, data augmentation, and ensemble inference can significantly enhance reasoning over long documents, achieving state-of-the-art results on Japanese document VQA benchmarks \cite{li2025adadocvqaadaptiveframeworklong}.

Evaluating code generation models remains challenging. Common metrics such as BLEU, CodeBLEU \cite{postCallClarityReporting2018}, and accuracy often fail to capture aspects like code readability, maintainability, and logical correctness. Additionally, existing datasets may not adequately represent the diversity and complexity of real-world programming scenarios, raising concerns about the generalizability and robustness of these models.

\subsection{Code Retrieval}
\textbf{Code Retrieval}, the task of finding relevant code snippets based on a query, is essential in applications like code recommendation, bug detection, and automated code completion. Early approaches relied on keyword-based search techniques, indexing code using tokens or syntactic features such as function names and variable identifiers. While effective for straightforward queries, these methods faltered when handling more complex searches that require an understanding of code semantics or context.

Deep learning has significantly advanced code retrieval by enabling the encoding of both code and natural language queries into continuous vector spaces. Models like CodeBERT \cite{fengCodeBERTPreTrainedModel2020}, GraphCodeBERT \cite{guoGraphCodeBERTPretrainingCode2021}, and CodeT5 \cite{wangCodeT5IdentifierawareUnified2021} utilize Transformer architectures to jointly model code and queries. Pretraining on large-scale code repositories allows these models to comprehend code semantics, facilitating more accurate retrieval based on natural language descriptions. Some researchers also focus on the important issues of position bias and other factors on model performance \cite{wang2025position, wang2025model,li2025taco}.

Graph-based models have further enhanced code retrieval by capturing structural dependencies inherent in code, such as data flow and control flow. GraphCodeBERT \cite{guoGraphCodeBERTPretrainingCode2021} incorporates Graph Neural Networks (GNNs) to represent code as graphs, where nodes denote entities like variables and functions, and edges represent their relationships. This representation enables the model to grasp finer-grained semantic information, improving retrieval accuracy for complex code queries.

\subsection{Retrieval-Augmented Generation (RAG)}
\textbf{Retrieval-Augmented Generation (RAG)} in code generation builds on the broader concept of retrieval-augmented learning in natural language processing \cite{lewisRetrievalAugmentedGenerationKnowledgeIntensive2020}. RAG models enhance code generation by dynamically sourcing relevant code snippets during the generation process, which is particularly beneficial in dynamic software development environments where specific libraries or frameworks may not be fully captured in the training data \cite{liu2024rag,liu2025qfft}.

Recent advancements in neural retrieval techniques, including dense retrievers and BM25-based methods, have improved the relevance and quality of retrieved code snippets \cite{zhouDocPromptingGeneratingCode2023}. Frameworks like RepoCoder \cite{zhangRepoCoderRepositoryLevelCode2023} achieve repository-level code completion by retrieving relevant functions across files and functions, enhancing the accuracy of code generation and better aligning with user intent. Similarly, RAMBO \cite{buiRAMBOEnhancingRAGbased2024} introduces strategies to identify repository-specific elements such as classes, methods, and variables, improving the semantic understanding of the retrieval module and optimizing code generation quality.

Benchmark studies like CodeRAG-Bench \cite{wangCodeRAGBenchCanRetrieval2024} evaluate the potential of retrieval augmentation in code generation across various datasets. However, these studies are often limited to predefined knowledge bases and do not fully explore applications in dynamic programming environments. Additionally, datasets like SWE-Bench \cite{jimenezSWEbenchCanLanguage2024} provide valuable real-world scenarios for code completion but lack explicit mappings between user pull requests and the functions to be modified, limiting their applicability in certain tasks. In contrast, RepoCoder’s multi-level approach—addressing line-level, function-level, and API-level tasks—offers targeted testing scenarios, though it still falls short of fully replicating the complexity of real-world programming contexts.

\section{RepoAlign-Bench Construction Implementation}

\subsection{Filtering Pipeline}

Our validation strategy systematically filters GitHub pull requests (PRs) to identify high-quality, non-trivial cross-component code changes through these sequential steps:

\begin{enumerate}
    \item \textbf{Source Aggregation}: Curate initial project pool from established benchmarks (SWE-Bench, Py150)
    
    \item \textbf{PR-Issue Linkage Verification}: Automatically verify each PR links to a corresponding issue using regex matching on commit messages and PR bodies for keywords like ``fixes \#issue-number'' or ``closes \#issue-number''
    
    \item \textbf{Static Analysis Quality Gate}: Apply PyLint framework to filter out trivial changes (whitespace, docstring updates, minor refactoring without logic changes)
    
    \item \textbf{Complexity-Based Filtering}: Analyze code diffs using cyclomatic complexity metrics. Calculate complexity changes in modified functions and exclude PRs below predefined threshold, ensuring meaningful logic/structural impact
\end{enumerate}

\subsection{AST Node Alignment Algorithm}

To link natural language change requests with specific code modifications, we align AST nodes with commit diffs using this process:

\begin{algorithm}
\caption{Aligning AST Nodes with Commit Diffs}
\begin{algorithmic}[1]
\State \textbf{Input:} Commit $C$, Repository States (before/after)
\State \textbf{Output:} Aligned pairs $P = \{(\text{query}, \text{modified\_code})\}$

\State Initialize $P \leftarrow \emptyset$
\State Get diff $D$ from commit $C$
\State Parse ASTs using Tree-sitter for both states
\State Extract query from PR description

\For{each modified file $F$ in $D$}
    \For{each hunk $H$ in $F$}
        \State Get line numbers (start, end)
        \State Map hunk to containing function/class in both ASTs
        \State Find nodes at specified lines
        \If{valid nodes found}
            \State Extract signature and body from after-state
            \State Add (query, modified\_code) to $P$
        \EndIf
    \EndFor
\EndFor
\State \textbf{return} $P$
\end{algorithmic}
\end{algorithm}

\subsection{Dependency Graph Screening}

\subsubsection*{Graph Construction}
Construct static call graphs for each repository version, where nodes represent functions/classes and directed edges represent dependencies (calls, inheritance, imports).

\subsubsection*{Pattern-Based Screening}
Analyze ``before'' and ``after'' dependency graphs to identify meaningful, non-local changes:

\begin{itemize}
    \item \textbf{API Propagation}: Function signature changes causing modifications in multiple downstream functions across different modules
    
    \item \textbf{Co-change Patterns}: Modifications in function groups not directly connected in call graphs but frequently changed together, indicating semantic coupling
    
    \item \textbf{Dependency Restructuring}: Addition/removal of dependency graph edges, indicating significant component interaction refactoring
\end{itemize}

\end{document}